\documentclass[12pt,letterpaper]{article}         
\pdfoutput=1
\usepackage{jheppub}

\def\be{\begin{eqnarray}}
\def\ee{\end{eqnarray}}

\newcommand\para{\paragraph{}}

\newcommand{\eqn}[1]{(\ref{#1})}

\def\Dslash{\,\,{\raise.15ex\hbox{/}\mkern-12mu D}}
\def\Dbarslash{\,\,{\raise.15ex\hbox{/}\mkern-12mu {\bar D}}}
\def\delslash{\,\,{\raise.15ex\hbox{/}\mkern-9mu \partial}}
\def\delbarslash{\,\,{\raise.15ex\hbox{/}\mkern-9mu {\bar\partial}}}
\def\pslash{\,\,{\raise.15ex\hbox{/}\mkern-9mu p}}
\def\calDslash{\,\,{\raise.15ex\hbox{/}\mkern-12mu {\cal D}}}

\def\lae{\mathrel{\mathop{\smash{\lower .5 ex \hbox{$\stackrel<\sim$}}}}}
\def\lae{\mathrel{\mathop{\smash{\lower .5 ex \hbox{$\stackrel>\sim$}}}}}

\preprint{DAMTP-2016-29}

\title{\large Universal Diffusion in Incoherent Black Holes}

\author{Mike Blake} 

\affiliation{ Department of Applied Mathematics and Theoretical Physics,
 University of Cambridge, Cambridge, CB3 OWA, UK}

\emailAdd{m.a.blake@damtp.cam.ac.uk}

\abstract{We study charge and energy diffusion in simple holographic theories with broken translational symmetry. We find that when the effects of momentum relaxation are very strong the diffusion constants take universal values $D_{c} \sim D_{e} \sim \hbar v_B^2/(k_B T)$. Here $v_B$ is the velocity of the butterfly effect and the coefficients of proportionality depend only on the scaling exponents of the infra-red fixed point. Our results suggest that diffusion in these incoherent black holes is controlled by $\tau \sim {\hbar}/(k_B T)$ independently of the mechanism of momentum relaxation. }

\begin{document}
\maketitle
\pagestyle{plain} \setcounter{page}{1}
\newcounter{bean}
\baselineskip16pt
\section{Introduction} 

\paragraph{}Many strongly correlated metals display a robust linear resistivity. It has long been suggested that such behaviour could be understood if transport properties were controlled by a universal dissipative timescale $\tau \sim {\hbar}/(k_B T)$ \cite{subirbook, sachdev, planckian}. In particular recent experiments have directly observed this same `Planckian' time-scale in the scattering rates of a wide range of materials exhibiting a linear resistivity \cite{mackenzie} . 
\paragraph{}Nevertheless, it remains unclear how such universality could occur across a range of materials whose microscopic physics and scattering mechanisms can differ massively. Inspired by an analogy with the viscosity bound of Kovtun, Son and Starinets  \cite{kss2}, \cite{incoherent} proposed that universal transport could arise from the saturation of a `Planckian' bound on the charge and energy diffusion constants
\be
D \sim \frac{\hbar v^2}{k_B T} 
\label{diff}
\ee
where $v$ is a characteristic velocity of the system. In a metal, one expects the characteristic velocity to be set by the Fermi energy. Since the diffusion constants are related to the conductivities via the Einstein relations, then the appeal of \eqn{diff} is that the saturation of such a bound would lead to a universal linear resistivity\footnote{Here we are assuming that the susceptibilities are temperature independent.}.
\paragraph{}A natural environment in which to expect a universal relationship such as \eqn{diff} to hold is in the charge diffusion constant of a particle-hole symmetric theory \cite{kovtun}. 
In this case the electrical conductivity decouples from momentum and hence is finite even in a translationally invariant theory. In \cite{butterflymike} we studied the charge diffusion constant of general holographic scaling theories and found a universal regime in which the diffusion constant was given by
\be
D_{c} = C \frac{\hbar v_B^2}{ k_B T}
\label{chargediff}
\ee
where $v_B$ is the velocity of the `butterfly effect' \cite{butterflyeffect,localisedshocks,stringyeffects,chaos,multiple,kitaev,roberts,cft,channels,eric} and $C$ is a constant that depended on the universality class of theory.
\paragraph{}In contrast the behaviour of energy diffusion, or diffusion in a metal where the electrical current couples to momentum, is more complicated. In a translationally invariant theory $D_{e}$ will diverge and hence it is extremely sensitive to the momentum relaxation rate $\Gamma$. Since this will depend on the precise way in which the translational symmetry is broken, one cannot expect to see universal behaviour in general.  However the proposal of \cite{incoherent} is that \eqn{diff} might apply to incoherent metals, where strong momentum relaxation is an intrinsic property of the theory.
\para{}Whilst there are very few theoretical approaches to incoherent metals, explicit examples of holographic models with strong momentum relaxation have been constructed \cite{vegh, blaise2,donos1,donos2,andrade}. The purpose of this paper is to study the diffusion constants of these theories. In order that charge and energy transport decouple, we will focus on incoherent theories with a particle-hole symmetry\footnote{We will briefly discuss how our ideas generalise to finite density in Section~\ref{discussion}.}. Since the electrical current decouples from momentum we find, just as in \cite{butterflymike}, that these models have a universal regime in which the charge diffusion constant is given by \eqn{chargediff}.
\paragraph{} As expected the behaviour of the energy diffusion constant is more complicated. When the translational symmetry is weakly broken, the energy diffusion constant is non-universal and depends on the momentum relaxation rate $\Gamma$. However when momentum relaxation is very strong the details of how we break the translational symmetry become unimportant. 
Rather, in this incoherent regime we find that the energy diffusion constant is always related to the butterfly effect as\footnote{Note that the diffusion constants in \eqn{chargediff} and \eqn{resultenergy} do themselves depend on the way translational symmetry is broken. However these effects solely reflect how the characteristic velocity, $v_B$, is changed due to the presence of momentum relaxation. }  
\be
D_{e} = E  \frac{\hbar v_B^2}{ k_B T}
\label{resultenergy}
\ee
where $E$ is another universal constant (which is different from $C$). Our results therefore support the suggestion that diffusion in incoherent metals
saturates \eqn{diff}. Indeed we find it striking how, now that we have identified a characteristic velocity $v_B$, these models precisely illustrate the proposal of \cite{incoherent}. 
\paragraph{}In Section~\ref{axionsection} we begin by studying the diffusion constants of a neutral black hole in which momentum relaxation is incorporated through linear sources for massless scalars \cite{andrade}. This provides a simple toy model in which we can illustrate how universality can emerge in the incoherent limit. In Section~\ref{qlattices} we consider a more general family of solutions known as `Q-lattices'  \cite{donos1,donos2,blaise2} and demonstrate that this universality continues to hold. Finally, we close with a discussion of our results in the context of \cite{mackenzie, incoherent}. 
\section{Linear Axions}
\label{axionsection}
\paragraph{}In order explain the basic ideas of this paper, we will start with the simplest holographic model of transport. This just consists of the Einstein-Maxwell action coupled to massless scalars\footnote{The diffusion constant of a similar massive gravity model were studied in the context of \cite{incoherent} in \cite{andrea}. However, without a definition of the characteristic velocity $v$ they were unable to see the universality we observe.}. In particular, if we work in four bulk dimensions we can consider the action
\be
S = \frac{1}{16 \pi G_N} \int \mathrm{d}^{4}{x}\sqrt{-g} \bigg [ R + \frac{6}{L^2} - \frac{1}{4 e^2} F^{\mu \nu} F_{\mu \nu} - \frac{1}{2} g^{\mu \nu} \partial_{\mu} \chi_{\cal A} \partial_{\nu} \chi_{\cal A} \bigg]
\label{axionaction}
\ee
where  ${\cal A} = 1, 2$ runs over the two spatial coordinates of the boundary quantum field theory. 
\paragraph{}As we mentioned in the introduction, we will restrict our attention to the case where there is no net charge density. The electrical current therefore decouples from momentum and hence the conductivity will be insensitive to the momentum relaxation rate. In contrast, to get a finite thermal conductivity we need to break the translational symmetry. 

\paragraph{}In order to do this we will introduce linear sources for the axion fields $\chi_{\cal A} = k x_{\cal A}$ that implement momentum relaxation in the boundary theory. Whilst the use of linear sources may appear unphysical, we will see in Section~\ref{qlattices} that these axion fields can more generally be viewed as the phase of a scalar `Q-lattice'. In this context such sources are therefore related to a periodic deformation of the boundary theory. 

\paragraph{}The advantage of using these linear sources is then that, even though they break translational symmetry, the background metric remains homogeneous. Indeed a black-hole solution can be written down analytically as \cite{andrade}
\begin{eqnarray} 
ds^2 = -\frac{r^2 U(r)}{L^2} dt^2 &+& \frac{L^2}{r^2 U(r)} dr^2 + \frac{r^2}{L^2}(dx^2 + dy^2) \nonumber \\
A_t = 0 \;\;\;\;\;\;\;\ \chi_1 &=& k x  \;\;\;\;\;\;\;\; \chi_2 = k y   
\label{axionmetric}
\end{eqnarray} %
where the emblackening factor $U(r)$ is given by
\be
U(r) = 1 - \frac{k^2 L^4}{2 r^2}  - \bigg({1 - \frac{k^2 L^4}{2 r_0^2}}\bigg) \frac{r_0^3}{r^3}
\ee
This black hole has a horizon at radius $r_0$ that determines the temperature, $T$ according to
\be
4 \pi T L^2= \bigg(3 r_0 - \frac{k^2 L^4}{2 r_0} \bigg)
\label{temperature}
\ee
and the entropy density is given by the Bekenstein-Hawking formula
\be
s = \frac{1}{4 G_N}\frac{r_0^2}{L^2}
\label{entropy}
\ee
The physics of the boundary theory can be described by a single dimensionless parameter $k/T$. When $k/T \ll 1$ then the translational symmetry is only weakly broken and the theory is described as
`coherent'. In this limit the background metric is well-approximated by the Schwarzchild solution and the only effect of the axions is to cause momentum to slowly relax \cite{davison}  at a rate $\Gamma
\sim k^2/T$. 
\paragraph{}Conversely once $k/T \gtrsim 1$ we have an `incoherent' metal in which momentum relaxation 
is a strong effect and cannot be treated perturbatively. In particular in this regime it is the axions themselves that are now responsible for sourcing the background
geometry. From \eqn{temperature} we can see that at low temperatures the horizon radius approaches a constant $r_0^2 = k^2 L^4/6$ and so the near-horizon
geometry will now correspond to an $AdS_2 \times R^2$ metric
\be
ds^2 = - \frac{3 {\tilde r}^2}{L^2} dt^2 + \frac{L^2}{3 {\tilde r}^2} d{\tilde r}^2 + \frac{k^2 L^2}{6} ( dx^2 + dy^2 )
\label{nearhorizon}
\ee
\subsection*{Charge Diffusion}
\paragraph{}The reason for starting with this simple model is that it is straightforward to write down analytic expressions for the transport coefficients and hence the diffusion constants. 
In particular, because we are dealing with the particle-hole symmetric theory, the electrical conductivity is just a constant
\be
\sigma = \frac{1}{e^2}
\ee
Whilst this is a simple formula, it is important to stress that the electrical conductivity itself is not a universal quantity - it explicitly depends on the normalisation
 of the current. To construct something independent of this normalisation, we can divide through by the charge susceptibility to obtain the diffusion constant
\be
D_{c} = \frac{\sigma}{\chi} \;\;\;\;\;\;\;\;\;\;\;\;\; \chi = \bigg( \frac{\partial \rho}{\partial \mu} \bigg)_{T}
\label{einsteincharge}
\ee
For our axion model we simply have $\chi^{-1} = e^2 {L^2}/{r_0}$ and hence the diffusion constant is given by 
\be
D_{c} = \frac{L^2}{r_0} 
\label{diffaxion}
\ee
\paragraph{}In order to write this diffusion constant in the form \eqn{diff} we first need to identify a characteristic velocity of our theory. As we argued in \cite{butterflymike}, one natural way to define such a velocity in a holographic theory is provided by the butterfly effect \cite{butterflyeffect, localisedshocks, roberts, multiple,kitaev,stringyeffects,chaos,cft,eric,channels}. In particular by studying this effect one can identify the characteristic speed, $v_B$, at which quantum information propagates in a strongly coupled theory. 
\paragraph{}Whilst the discussion in \cite{butterflymike} centred around translationally invariant theories, this velocity is only sensitive to the background geometry.
 It is therefore straightforward to apply the the shock-wave techniques of \cite{localisedshocks, stringyeffects, butterflymike, roberts} to calculate this velocity for our homogeneous metric.  In particular for any metric of the form \eqn{axionmetric} this velocity is given by (see Appendix~\ref{appendixa}) 
\be
v_B^2 = \frac{\pi T L^2}{ r_0} 
\ee 
where the effects of momentum relaxation are implicitly contained in the dependence of the horizon radius on the ratio $k/T$\footnote{Note that since $v_B$ is sensitive only to the background geometry information can still spread ballistically even in theories with momentum relaxation. Intuitively information can be carried by degrees of freedom, such as `particle-hole pairs', that do not carry any net momentum.}. We therefore see that the diffusion constant of our axion models can be written as
\be
D_{c} = \frac{v_B^2}{\pi T} 
\label{diffcharge}
\ee
which holds independently of any of the parameters $e, L, k, T$ in our bulk theory. 
\paragraph{} It is worth emphasising that whilst the conductivity is just a constant, $D_{c}$ itself actually depends on the strength of momentum relaxation $k/T$ through $r_0$.  
In fact, the dimensionless diffusion constant $D_{c} T$ vanishes in the incoherent limit, and hence it was suggested in \cite{andrea} that it is not possible to formulate a bound on the diffusion constants for models such as \eqn{axionaction}.
\paragraph{} However, we can now see that this dependence simply reflects the fact that turning on sources for the axion fields will change the characteristic velocity, $v_B$, of the theory. Having identified this velocity we see that the relationship \eqn{diffcharge} always holds and as such the 
timescale we would extract from the diffusion constant as $D_{c} \sim v_B^2 \tau$ is consistent with a Planckian bound \eqn{diff}
\be
\tau \sim \frac{1}{2 \pi T}
\label{tau}
\ee
\subsection*{Energy Diffusion}
\paragraph{}In contrast to the electrical conductivity, energy transport in our theory is sensitive to the details of momentum relaxation. Indeed for this
axionic model the thermal conductivity, ${\kappa}$, is given by \cite{donos3}
\be
{\kappa} = \frac{4 \pi s T}{k^2}
\label{thermal}
\ee
and hence diverges in the translationally invariant theory. In order to calculate the diffusion constant, we can once again make use of an Einstein relation to extract $D_{e}$ from the thermal conductivity and the specific heat $c_{\rho}$
\be
D_{e} = \frac{{\kappa}}{c_\rho} \;\;\;\;\;\;\;\;\ c_{\rho} = T \bigg( \frac{\partial s}{\partial T} \bigg)
\label{einsteinthermo}
\ee
which yields \cite{davison}
\be
 D_{e} = \frac{1}{2 k^2 L^2} \bigg( 3 r_0+ \frac{k^2 L^4}{2 r_0} \bigg)
\ee
We can now explicitly see that, as discussed in the introduction, it is not in general possible to write the energy diffusion constant in the same form as $D_{c}$ \eqn{chargediff}. In particular, when momentum relaxation is weak the diffusion constant exhibits the expected divergence
\be
D_{e} \sim T/k^2  \;\;\;\;\;\;\;\;\;\;\;\;\;\;\;\;\;\;\;  k/T \ll 1
\ee 
and hence its value is set by the momentum relaxation rate $\Gamma \sim k^2/T$. Since this depends on how strongly the translational symmetry is broken, it is highly non-universal\footnote{At high temperatures the characteristic velocity $v_B$ is just a constant.}. 
\paragraph{}However the proposal of \cite{incoherent} is that universality could emerge in the incoherent limit. Indeed in this limit we see that the diffusion constant just approaches 
\be
D_{e} = \frac{L^2}{2 r_0}  \;\;\;\;\;\;\;\;\;\;\;\;\;\;\;\;\;\;\;  k/T \gg 1
\ee
which, up to an order one number, is precisely the same value that the charge diffusion constant took. Therefore whilst the energy diffusion constant still depends on $k/T$, this is again solely due to the fact that the characteristic velocity $v_B$ is changing. In other words, in the incoherent regime the energy diffusion constant of our axion model can be written as
\be
D_{e} = \frac{v_B^2}{2 \pi T}
\ee
and hence is governed by the same universal timescale
\be
\tau \sim \frac{1}{2 \pi T}
\ee
that controlled charge transport. 
\begin{figure}
\begin{center}
\resizebox{75mm}{!}{\includegraphics{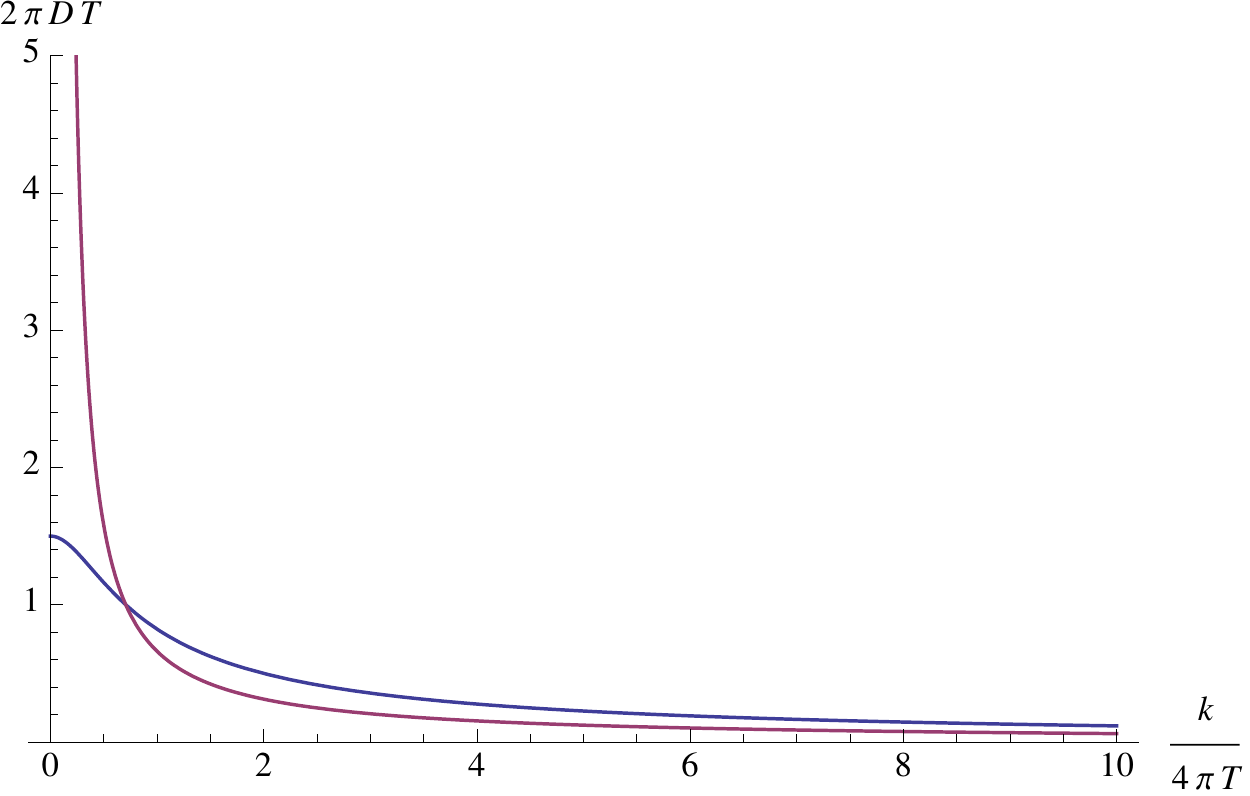}}
\resizebox{75mm}{!}{\includegraphics{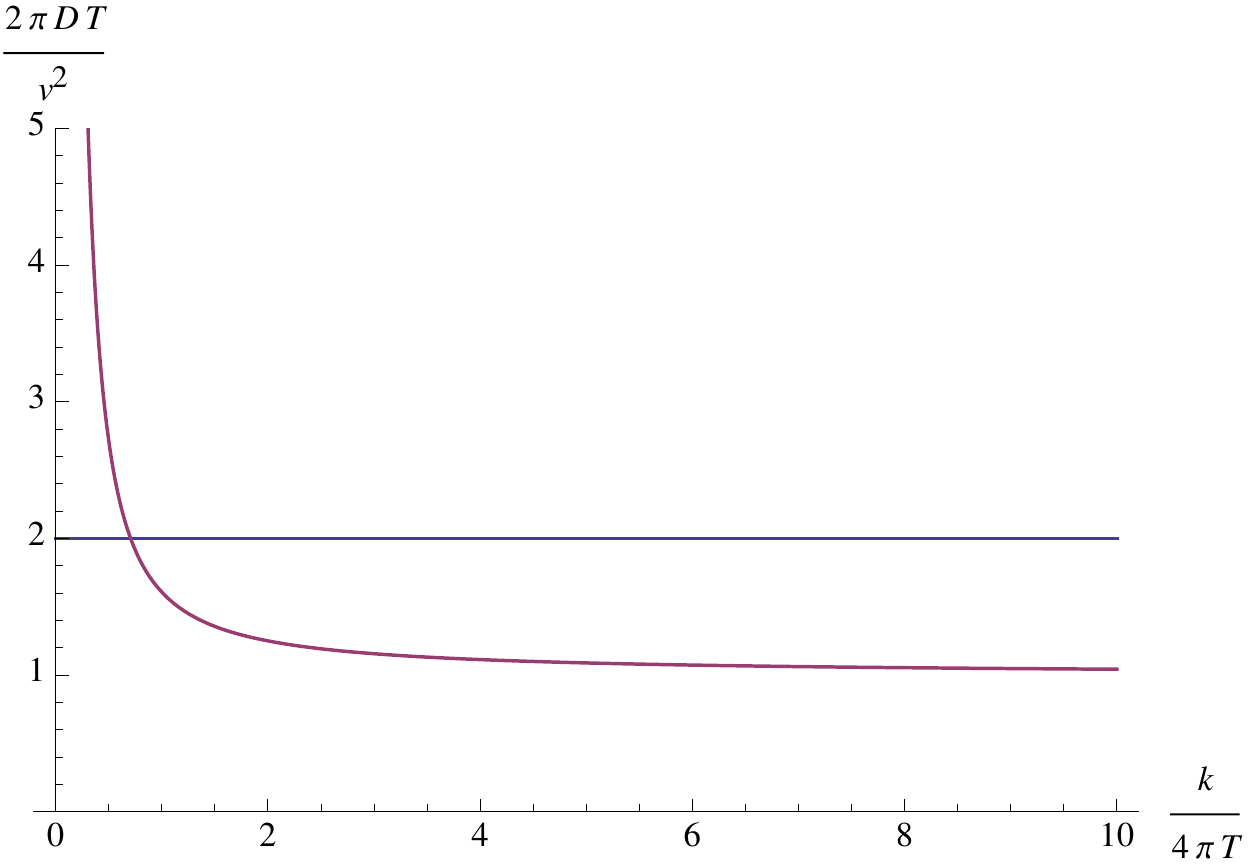}}
\end{center}
\label{neutralratio}
\caption{Diffusion constants of neutral axion model. The diffusion constants themselves (left) are sensitive to the ratio $k/T$. However, upon dividing by the butterfly velocity $v_B^2$ (right) the universality is evident. In particular note that the energy diffusion constant (red) becomes universal at large $k/T$. The diffusion constants cross at the self-dual point \cite{davison}
$k L^2 = \sqrt{2} r_0$.}
\end{figure} 
\paragraph{}In order to illustrate this behaviour we have plotted the diffusion constants in Figure 1. The left hand figure shows the diffusion constants themselves which are non-universal and vanish
in the limit $k/T \gg 1$. On the right hand side we have divided through by the butterfly velocity to construct the ratios $2 \pi D T/v_B^2$. In our simple geometry $D_{c}$
is just a constant in these units. In contrast the energy diffusion constant is more complicated, and can be seen to diverge in the translationally invariant limit. 
\paragraph{}However, as we increase the strength of momentum relaxation then we find that $2 \pi D_e T/v_B^2$ cannot be made arbitrarily small but instead saturates
at an ${\cal O}(1)$ value. As soon as we reach the incoherent regime $k \sim 4 \pi T$ then we will be close to this saturating value and hence the diffusion constants
will be universally given by
 \be
 D_{c} = \frac{v_B^2}{\pi T} \;\;\;\;\;\;\;\;\;\;\;\;\;\;\;   D_{e} \approx \frac{v_B^2}{2 \pi T}
 \label{diffscaling}
 \ee
Precisely this phenomenology was proposed in \cite{incoherent} in the context of a fundamental bound on the diffusion constant. We find it remarkable that the simplest holographic model of an incoherent metal realises this behaviour. One might worry, however, that these results are an artefact of our choice of action \eqn{axionaction}. In the next section we will therefore consider a much wider class of incoherent holographic geometries. We will find that the universality we see in the diffusion constants of this axion model continues to hold more generally. 
\section{Q-Lattice Models}
\label{qlattices}
\paragraph{}We have just seen that in the incoherent limit $k \gg T$ both the charge and energy diffusion constants of the axion model were universal. That is we had $D  T \sim v_B^2$ independently of the strength of momentum relaxation. In this section we want to show that such behaviour occurs more generally in holographic models of incoherent metals.
\paragraph{}In particular we wish to consider the so-called holographic `Q-lattice models' that were introduced in \cite{donos1}. These models consist of coupling the axion model
of the last section to a dilaton field 
\begin{eqnarray}
S = \int \mathrm{d}^4x \sqrt{-g} \bigg [ R - \frac{c}{2} ( (\partial \phi)^2 + Y(\phi) ((\partial \chi_1)^2 + (\partial \chi_2)^2)  -  V(\phi)  - \frac{1}{4} Z(\phi) F^{\mu \nu} F_{\mu \nu} \bigg] \nonumber \\
\label{qlattice}
\end{eqnarray}
Roughly speaking, one can think of these models as arising from decomposing a complex scalar field into its magnitude and phase as $\Psi \sim \phi e^{i \chi }$. Once again the solutions we are interested correspond to breaking translational symmetry by turning on a linear source for the axions $\chi_{\cal A} = k x_{A}$. The name `Q-lattice' then reflects the fact that, in terms of the original complex scalar $\Psi$ this appears to be a periodic deformation of the boundary theory.  
\paragraph{}By choosing different actions for $\Psi$ one can engineer different potentials for the dilaton field $\phi$. Here, we will assume that
when the dilaton becomes large the leading form of these potentials is an exponential 
\be
Y(\phi) = e^{2 \phi} \;\;\;\;\;\;\;\ V(\phi) = - V_0 e^{\alpha \phi} \;\;\;\;\;\; Z(\phi) = Z_0^2 e^{\gamma \phi}
\ee
Note that the parameter $c$ cannot be set to one without changing these exponents, and so it is important in determining the form of the solutions. Since the scalar potential is unbounded, then this action can be used to construct solutions where the size of the lattice diverges $\phi \rightarrow \infty$ in the infra-red \cite{donos2,blaise2}. These solutions therefore describe incoherent metals, in which the effects of the lattice are becoming extremely strong at low temperatures. 
\subsection*{Incoherent Scaling Geometries}
\paragraph{} Just like in the axion model, the resulting metrics are homogeneous. They correspond to the well-studied class of metrics known as hyperscaling violating geometries \cite{dong,kachru,strangemetals,huijse,elias1,elias2}. In particular there exist a family of neutral solutions of the form\footnote{Note our radial coordinate differs from the more familiar one in \cite{dong} by $r = {\tilde r}^{\theta-z}$.}
\begin{eqnarray} 
ds^2 = - U(r) dt^2 &+& \frac{dr^2}{U(r)} + V(r)(dx^2 + dy^2)  \nonumber \\
A_t = 0 \;\;\;\;\;\;\;\ \chi_1 &=& k x  \;\;\;\;\;\;\;\; \chi_2 = k y   
\label{metricgeneral}
\end{eqnarray}
where the metric functions are given by
\be
U(r)= L_t^{-2} r^{u_1} \;\;\;\;\;\;\;\;\; V(r) &=& L_x^{-2} r^{2 v_1} \;\;\;\;\;\;\;\; e^{\phi(r)} = e^{ \phi_0} r^{ \phi_1} \nonumber 
\label{ansatz}
\ee
The powers in our geometry are then related to a dynamical critical exponent, $z<0$, and a hyperscaling violation exponent $\theta>2$\footnote{These restrictions on the exponents are necessary to order to have a consistent geometry in which the scalar field diverges in the infra-red. In terms of our exponents this corresponds to the regime $2 \leq \alpha < \sqrt{4 + c}$.} as
%
%
\be
u_1= \frac{2 z - \theta}{z - \theta} \;\;\;\;\;\;\;\ 2 v_1 = \frac{2 - \theta}{z - \theta} \;\;\;\;\;\;\;  2 \phi_1 = \frac{2}{z - \theta}
\ee
and the exponents are determined by the choice of parameters $c, \alpha$
\be
z = \frac{4 - \alpha^2 + c}{4 -2 \alpha} \;\;\;\;\;\;\;\;\;\;\;\;\;\;  \theta = \alpha
\ee
Additionally, the Einstein equations tell us how the scales of the metric $L_t, L_x$ are generated by the Q-lattice
\begin{eqnarray}
 (z - \theta)^2 V_0 e^{\alpha \phi_0}  &=& L_t^{-2} (2 z - \theta) (2 + z - \theta)  \nonumber \\
(2 z -\theta)  c k^2 e^{2 \phi_0} &=& L_x^{-2} V_0 e^{\alpha \phi_0}(2 z- 2) 
\label{constraint}
\end{eqnarray}
\paragraph{} Whilst these metrics appear quite complicated, the main point is that they correspond to generalisations of the near-horizon geometry we saw in the Section~\ref{axionsection}. In particular, for the case our axion model $z = \infty, V_0 = 6/L^2, \phi =0, c =1$ these constraints reduce to
\be
 u_1 = 2 \;\;\;\;\;\;  v_1 = 0 \;\;\;\; L_t^{-2} = \frac{3}{L^2} \;\;\;\; L_x^{-2} =  \frac{k^2 L^2}{6} . 
\ee
and so we reproduce the metric \eqn{nearhorizon} we studied in the last section. However rather than just the simple $AdS_2\times R^2$ geometry, the Q-lattices can now support far more general scaling geometries parameterised by $(z, \theta)$.   
\paragraph{}Finally, in order to calculate the diffusion constants, we need to heat these solutions up to a finite temperature. We can do this by turning on an emblackening factor in our ansatz \eqn{ansatz}
\be
U(r)= L_t^{-2} r^{u_1}\bigg(1 - \frac{r_0^{\delta}}{r^{\delta}} \bigg)
\ee
where we have $\delta = u_1 + 2v_1-1$. It is simple to check that this deformation can be turned on without changing the rest of our bulk solution and corresponds to a temperature $4 \pi T = U'(r_0)$ for the boundary field theory. 
\subsection*{Charge Diffusion}
\paragraph{} Before focusing on energy diffusion, let us begin by studying the diffusion of charge in these theories. In \cite{butterflymike} we discussed charge diffusion for general
holographic scaling geometries with a particle-hole symmetry. Since the charge diffusion constant is only sensitive to the background metric, our analysis can also be applied to these Q-lattice models as well. 
\paragraph{}To extract the diffusion constant we will again use the Einstein relation \eqn{einsteincharge}. The conductivity of a dilaton model just corresponds to the effective Maxwell coupling at the horizon. This is now no longer a constant, but rather has a non-trivial temperature dependence
\be
\sigma = Z(\phi)|_{r_0} \sim T^{{(2 \Phi - \theta})/z}
\ee
where $\Phi$ is an anomalous scaling dimension for the charge density that arises due to the coupling, $\gamma$, between the gauge field and the dilaton \cite{blaise1,blaise2,andreas1,andreas2}
\be
\gamma \phi_1 = \frac{2 \Phi - \theta}{z - \theta}
\ee
\paragraph{}Although the conductivity of these dilaton models is now more complicated, the diffusion constant can still take a simple form. This is because the charge susceptibility is also sensitive to the profile of the dilaton \cite{kss1,iqballiu}
\be
\;\;\;\;\;\;\;\;\; \chi^{-1} = \int_{\infty}^{r_0} \mathrm{d}{r}   \frac{1}{\sqrt{-g} Z(\phi) g^{rr} g^{tt} }
\label{susc}
\ee
and hence the effects of the running Maxwell coupling can effectively cancel.
\paragraph{} In particular, the behaviour of diffusion constant will depend on which region of the geometry dominates the integral in \eqn{susc}. In order to characterise this, it is useful to introduce the scaling dimension, $\Delta_{\chi} =  2 - \theta + 2 \Phi - z $ of the susceptibility. Now since the contribution to 
$\chi^{-1}$ from near the horizon scales like $ T^{-\Delta_{\chi}/z}$ then for low temperatures the infra-red region of the geometry will dominate the integral whenever $\Delta_{\chi}/z > 0$. 
\paragraph{} Since diffusion is then controlled by the near-horizon physics, it is natural to expect a connection with the butterfly effect. Upon evaluating the integral
 \eqn{susc} we have that the diffusion constant is related to the horizon radius by 
\be
D_{c} = \frac{z - \theta}{\Delta_{\chi}} L_x^2 r_0^{1 - 2 v_1}
\ee
To compare with \eqn{diff} we need the characteristic velocity of our theory. As was shown in \cite{butterflymike,roberts}, and we review in Appendix~\ref{appendixa}, the
butterfly velocity for a general metric of the form \eqn{metricgeneral} is 
\be
v_B^2 = \frac{ 2 \pi T}{V'(r_0)} = \frac{2 \pi T}{2 v_1} L_x^2 r_0^{1 - 2 v_1} 
\label{butterflylattice1}
\ee
and so we see that the diffusion constant is universally given by 
\be
D_{c} = \frac{2 - \theta}{\Delta_{\chi}} \frac{v_B^2}{2 \pi T}
\label{diffcharge2}
\ee
\paragraph{} It is worth stressing that, just as in the the axion model, both the diffusion constant and the butterfly velocity depend through $L_x$ and $r_0$ on the details of the Q-lattice solution (and in particular on the strength of momentum relaxation). However provided we are in this universal regime\footnote{ On the other hand when $\Delta_{\chi}/z < 0$ it will be the UV region of the geometry which dominates the integral. We therefore cannot calculate the diffusion constant just from knowledge of our infra-red scaling theory. The diffusion constant is then no longer related to the butterfly effect in a universal manner, but rather it will be parametrically larger than $v_B^2/T$ by powers of the UV cutoff \cite{butterflymike}.} $\Delta_{\chi}/z > 0$ then we see that the relationship between them is always given by \eqn{diffcharge2}. As such in all these different theories we will have that charge diffusion is universally controlled by the same Planckian timescale $\tau \sim 1/T$ that we saw in the axion model. 
\subsection*{Energy Diffusion}
\paragraph{}We can now turn to the question of energy diffusion in these Q-lattice models. Recall that in the axion model we saw that the effects of the axion fields
on the geometry were precisely such that the diffusion constant became universally related to $v_B$ in the incoherent limit. Our goal is to understand whether the same
thing happens for our more general metrics \eqn{ansatz}.  For the final time, we therefore invoke the Einstein relation \eqn{einsteinthermo} to compute $D_{e}$ from 
knowledge of the thermal conductivity $\kappa$ and the specific heat $c_{\rho}$.
\paragraph{} For these holographic Q-lattice models, it is now well known that one can obtain an analytic expression for the thermal conductivity 
in terms of properties of the black hole horizon \cite{donos3}. In particular this formula relates the thermal conductivity to the size of the 
lattice at the horizon according to\footnote{It is worth noting that the thermal conductivity bound of \cite{saso2} does not apply to models with an unbounded potential, hence why $\kappa/T$ can vanish at low temperatures. Nevertheless the diffusion constant will still be universal and hence it would provide a more natural object than the conductivity on which to formulate rigorous `bounds'.}
\be
\kappa = \frac{4 \pi s T }{c k^2 e^{2 \phi}} \bigg|_{r_0} \sim T^{(z-\theta)/z}
\ee
If we extract the specific heat from the scaling of the entropy density 
\be
s \sim T^{(2 - \theta)/z} 
\ee
then we can deduce that the diffusion constant must be given by
\be
D_{e} =  \frac{4 \pi z T}{(2 - \theta) c k^2e^{2 \phi}} \bigg|_{r_0}
\label{difflattice}
\ee
\paragraph{}Upon comparing this expression with the butterfly velocity \eqn{butterflylattice1} then it is clear that for an arbitrary bulk metric, there will not be a simple relationship between the energy diffusion
constant and $v_B$. This just reflects the fact that, as we explicitly saw in the axion model, we should not always expect to see universal behaviour in the diffusion constant. 
\paragraph{}However, in our incoherent theories, the metric is not some arbitrary function. Remember the key point is that the lattice itself 
is now responsible for creating the geometry. As a result, the Einstein equations relate the profile
of the scalar field to the metric and hence can be used to relate the diffusion constant \eqn{difflattice} to the butterfly effect. 
\paragraph{}In particular we can note that second equation in \eqn{constraint} tells us the length-scale $L_x$ in our scaling solution is not arbitrary, 
but rather is fixed in terms of the lattice profile $k^2 e^{2 \phi_0}$. In the axion model, this condition is equivalent to our observation that the axions source 
a ground state entropy. In these more general scaling theories we can again use this equation to relate the lattice profile to  the metric function $V(r)$ and hence the butterfly effect. 
\paragraph{}Indeed, after using the equations \eqn{constraint} we find that the diffusion constant of our scaling theories can be rewritten as\footnote{Note that the condition that the axion fields remain present in the equations of motion imposes the constraint $u_1 + 2 v_1 - 2 \phi_1 = 2$ which we have used in simplifying \eqn{difflattice}.}
\be
D_{e} = \frac{ z (z - \theta)}{(2 z - 2)(2 - \theta)} L_x^2  r_0^{1 - 2 v_1} 
\ee
which is exactly the same combination of parameters $L_x, r_0$ that appeared in the charge diffusion constant. Comparing with \eqn{butterflylattice1} we therefore see that the energy diffusion constant of these models is universally related to the butterfly effect by
\be
D_{e}  = \frac{z}{2 z - 2} \frac{v_B^2}{2 \pi T}
\label{latticeenergy}
\ee
and hence we extract the same timescale $\tau \sim 1/T$ independently of our choice of Q-lattice model. 
\paragraph{} In order to emphasise why this result is so surprising, it is instructive to again recall what happened when momentum relaxation was a weak effect. As we explicitly saw in the axion model, in such a case the energy diffusion constant is not related to $v_B$ in any simple manner, but rather depends on the details of momentum relaxation. However what we are seeing is that when momentum relaxation is a strong effect this dependence goes away. The key point in that in these incoherent theories it is now the Q-lattice itself that is responsible for supporting the scaling geometry. The Einstein equations then imply that whatever type of Q-lattice we turn on the resulting geometry is always such that the diffusion constant and the butterfly effect of our models are related by \eqn{latticeenergy}.
\section{Discussion}
\label{discussion}
\paragraph{} In this paper we have studied diffusion in simple holographic theories with broken translational symmetry. In particular we found that both the charge and energy diffusion constants of these models could be universally related to the butterfly velocity
\be
D_{c} = C \frac{\hbar v_B^2}{k_B T}      \;\;\;\;\;\;\;\;\;\;\;\;\;\;\;\;\; D_{e} = E \frac{\hbar v_B^2}{k_B T}
\label{bound}
\ee
where the constants of proportionality depended only on the scaling exponents of our theories. For the charge diffusion constant, the relationship \eqn{bound} held regardless of the strength of momentum relaxation. Conversely, for the energy diffusion constant it was necessary to be in the incoherent regime where the lattice itself supports the background geometry. 
\paragraph{} Since we have been considering theories with a particle-hole symmetry, the universality  of the charge diffusion constant could have been anticipated from the results of \cite{butterflymike} . In contrast, one would expect that energy diffusion should be highly sensitive to the way in which translational symmetry is broken. However we saw that in the incoherent limit this dependence went away,  and the energy diffusion constant also became universally tied to the butterfly velocity.  Heuristically, it seems that once we reach the Planckian rate we can no 
 longer increase the rate of dissipation by breaking the translational symmetry any further\footnote{Since the coefficients in \eqn{bound} depend on the universality class, and can be arbitrary small, it will not be possible to formulate a strict bound using $v_B$. Nevertheless the spirit that the diffusion constants are generically set by $v_B^2/T$ remains.}.  The microscopic details of momentum relaxation are therefore unimportant and the result is universality.
\paragraph{} Throughout this paper we focused for technical reasons on theories with a particle-hole symmetry. Now that we understand that in an incoherent theory both charge and energy diffusion can be universally governed by \eqn{bound} then it should be possible to extend our considerations to finite density. In this situation, the energy and charge currents overlap and the result is a pair of coupled diffusion equations with eigenvalues $D_{\pm}$. So long as we are in the incoherent regime, i.e. that the dominant degrees of freedom sourcing the geometry are the Q-lattice fields, then we expect that these diffusion constants will continue to take similar values as to\footnote{Note however that when the charge density becomes sufficiently strong to dominate the geometry we would expect to move away from the universal regime. This suggests that incoherent metals saturating \eqn{bound} have an approximate particle-hole symmetry in the infrared.} \eqn{bound}.
\paragraph{} These results therefore lend support to the proposal of \cite{incoherent} that diffusion in incoherent metals can be universal. To reiterate, our central observation is that even when momentum relaxation is strong the diffusion constants of our models could not be made parametrically smaller than the butterfly velocity. That is in the incoherent regime they were given by 
\be
D \sim \frac{\hbar v_{B}^2 }{k_B T}
\label{mikeproposal}
\ee
independently of the details of the theory or the strength of momentum relaxation. These holographic models therefore provide concrete examples of how universal transport properties governed by $\tau \sim {\hbar}/{k_ B T}$ could emerge in an incoherent metal \cite{incoherent, mackenzie}. 
\paragraph{}An important question for future work is to understand to what extent \eqn{mikeproposal} holds outside the simple class of theories studied here and in \cite{butterflymike}. For instance, one can consider more general holographic models in which the metric is not homogeneous. Recently \cite{steinberg} studied the connection between the charge diffusion constant and the butterfly velocity of such theories by  considering a hydrodynamic treatment of striped inhomogeneities. Interestingly they found that the diffusion constant continues to obey the scaling $D_{c} \sim v_B^2/T$ expected from our proposal \eqn{mikeproposal}, albeit with a coefficient of proportionality that is no longer universal\footnote{In particular the constant of proportionality depends on the profile of the inhomogeneities. It would be interesting to understand if there is another velocity, which differs only by a numerical factor from $v_B$, in terms of which this relationship can be made more universal.}. It would be worthwhile to perform a more detailed analysis of this inhomogeneous setting, and in particular to address the question of energy diffusion in the presence of strong disorder. 
\paragraph{}Finally, we note that it would be of great interest to investigate whether \eqn{mikeproposal} also holds in non-holographic theories. In particular \cite{stanford} has recently proposed a generalised Sachdev-Ye-Kitaev model that provides a soluble quantum mechanical model of an incoherent metal. Remarkably the energy diffusion constant and butterfly velocity of this model were found to obey a simple relationship $D_{e} = \hbar v_B^2/(2 \pi k_B T)$ in consistency with our proposal\footnote{Intriguingly, this is precisely the same as the result \eqn{diffscaling} for the axion theory we studied in Section~\ref{axionsection}.}. The results of \cite{stanford} therefore suggest a more general validity of our proposal, and provide an exciting new context in which to explore it further. 
\acknowledgments{I am grateful to Aristomenis Donos, Sean Hartnoll, Jorge Santos and David Tong for many useful conversations and comments on a draft. I also thank Brian Swingle and Dan Roberts for particularly thought-provoking discussions. This research was funded through a Junior Research Fellowship from Churchill College, University of Cambridge, and supported in part by the European Research Council under the European Union's Seventh Framework Programme (FP7/2007-2013), ERC Grant agreement STG 279943, Strongly Coupled Systems}
\appendix
\section{The Butterfly Effect}
\label{appendixa}
\paragraph{}A detailed discussion of the butterfly effect and the connection to quantum chaos can be found in \cite{localisedshocks,stringyeffects,multiple, butterflyeffect,chaos}. In the interests of making this paper self-contained, we will simply review how to extract the butterfly velocity $v_B$ for general metrics of the form \eqn{metricgeneral} using the shock-wave techniques of \cite{localisedshocks, stringyeffects}.  
\paragraph{}The butterfly effect is associated with the exponential growth of a small perturbation to a quantum system. In holography this effect has a beautiful realisation in terms of a gravitational shock-wave at the horizon of a black-hole \cite{butterflyeffect}. In particular, if one considers releasing a particle from the boundary at a time $t_w$ in the past, then for late times $t_w > \beta$ the energy density of this particle is localised near the horizon. In Kruskal coordinates $(u,v)$ the resulting stress tensor of this particle is then given by 
\be
\delta T_{u u} \sim E e^{\frac{2 \pi}{\beta} t_w} \delta(u) \delta(\vec{x})
\label{stresstensor0}
\ee
where $E$ is the asymptotic energy of the particle.   
\paragraph{}Due to the exponential boosting of the energy density the back reaction of this stress tensor eventually becomes significant and results in the formation of a shock-wave geometry \cite{butterflymike, localisedshocks,stringyeffects,butterflyeffect,chaos,multiple,roberts}. To construct these solutions we first need to write our metrics in Kruskal coordinates as
\be
ds^2 &=&  A( u v) du dv + V( u v) ( dx^2 + dy^2) 
\ee
Then the shock-wave corresponds to a solution where there is a shift in the $v$ coordinate $v \rightarrow v +  h(x)$ as one crosses the $u=0$ horizon. Such a metric can be described by
an ansatz
\be
ds^2 = A( u v) du dv +  V( u v ) d\vec{x}^2 - A(u v) \delta(u) h(x) du^2
\ee
where for quite general theories of Einstein gravity coupled to matter one finds a solution to the Einstein equations provided the shift obeys a Poisson equation \cite{sfetsos} 
\be
(\partial_{i}\partial_{i}  - m^2 ) h(x) \sim \frac{1 6 \pi G_N V(0)}{A(0)} E e^{\frac{2 \pi}{\beta} t_w} \delta(\vec{x})
\label{poisson}
\ee
with a screening length 
\be
m^2 = \frac{2}{A(0)} \frac{\partial{V(u v)}}{\partial{ (u v) }}\bigg|_{u=0}
\label{musquare}
\ee
In particular, as is discussed in detail in \cite{sfetsos, roberts}, one finds that after using the background equations of motion \eqn{poisson} and \eqn{musquare} continue to hold
even when there are non-trivial matter fields supporting the background geometry\footnote{Note that if there is a non-zero background stress tensor then in these coordinates there is a shift in the components of the stress tensor in addition to the change in the metric \cite{sfetsos,roberts}.}. As such the only way the matter content of the theory effects the shock wave is
through the determination of the background metric $A( u v)$, $V(u v)$. The net result is that, even though we have introduced scalar fields to break the translational symmetry, 
we can still apply these equations to construct the shock wave solutions for our homogeneous metrics \eqn{metricgeneral}. 
\paragraph{}It is then straightforward to solve \eqn{poisson}. At large distances one finds that the shift is given by 
\be
h(x) \sim \frac{ E e^{\frac{2 \pi}{\beta} (t_w - t_*)- m |x|}}{|x|^{\frac{1}{2}}}
\ee
where $t_{*} \sim \beta \mathrm{log} N^2$ is the scrambling time \cite{scrambling}. From the form of this solution we can then read off that the effects of the particle grow with a
Lyapunov exponent $\lambda_L = 2 \pi/\beta$ and propagate at the butterfly velocity $v_B = 2 \pi/ (\beta m)$. Upon swapping back to a radial coordinate system 
\eqn{metricgeneral} we now arrive at the formula for the butterfly velocity that we quoted in the main text
\be
v_B^2 = \frac{2 \pi T}{V'(r_0)}
\ee
\paragraph{} For the case of the axion model we have $V(r) = L^{-2} r^2$ and hence the butterfly velocity is given by
\be
v_B^2 = \frac{\pi T L^2}{r_0}
\ee
When momentum relaxation is weak we reproduce the Schwarzchild value $v_B^2 = 3/4$, whilst in the incoherent regime this
velocity vanishes as $v_B^2 \sim T/k$. 
\paragraph{} Finally for the Q-lattice solutions we have $V(r) = L_x^{-2} r^{2 v_1}$ and so the velocity is now
\begin{eqnarray}
v_B^2 &=& \frac{2 \pi T}{2 v_1} L_x^{2} r_0^{1 - 2 v_1} 
\end{eqnarray}
which results in the usual scaling $v_B^2 \sim T^{2 - 2/z}$ found in hyperscaling violating geometries \cite{butterflymike,roberts}.

\end{document}